\documentclass[preprint,aps,tightenlines]{revtex4}

\begin{document}



 
\title{Spontaneous Scale Symmetry Breaking in 2+1-Dimensional QED at
 Both Zero and Finite Temperature}
\author{M.E. Carrington}
\email{meg@theory.uwinnipeg.ca}
\altaffiliation{Winnipeg Institute for
Theoretical Physics, Winnipeg, Manitoba, R3B 2E9 Canada}
\affiliation{Department of Physics, 
Brandon University, Brandon, Manitoba,
R7A 6A9 Canada}
\author{W.F. Chen}
\email{wchen@theory.uwinnipeg.ca}
\altaffiliation{ Winnipeg Institute for
Theoretical Physics, Winnipeg, Manitoba, R3B 2E9 Canada}
\author{R. Kobes}
\email{randy@theory.uwinnipeg.ca}
\altaffiliation{ Winnipeg Institute for
Theoretical Physics, Winnipeg, Manitoba, R3B 2E9 Canada}
\affiliation{Department of Physics, University of Winnipeg, Winnipeg, 
Manitoba, R3B 2E9 Canada}

\begin{abstract}
A complete analysis of dynamical scale symmetry breaking 
in $2+1$-dimensional QED at both zero and finite temperature is presented
by looking at solutions to 
the Schwinger-Dyson equation. In different kinetic energy regimes 
we use various numerical and analytic techniques (including an expansion 
in large flavour number). It is confirmed that, contrary to the case of 
$3+1$ dimensions, there is no dynamical scale symmetry breaking at 
zero temperature, despite the fact that chiral symmetry breaking can 
occur dynamically.  At finite temperature, 
such breaking of scale symmetry may take place.
\vspace{3ex}
\end{abstract}


\maketitle

\section{Introduction}

Scale symmetry cannot be an exact symmetry in elementary particle
physics, since it would will enforce all the observed particles to be massless or 
to have a continuous mass spectrum \cite{ref1}. This explicitly contradicts 
experimental observation. Therefore,  scale invariance must be broken. 

In general, there are two kinds of scale symmetry breaking mechanisms. 
One is explicit breaking which occurs when dimensional parameters are 
present in the classical action. The other is anomalous 
scale symmetry breaking which happens at quantum level due to the necessity of 
implementing renormalization and the consequent occurrence of 
dimensional transmutation: a new momentum scale automatically arises.
In fact, this new scale parameter should be regarded as one of the most 
important elements underlying a quantum
field theory, since in a quantum field theory
it is the first derivative of the interaction coupling with 
 respect to the scale parameter,  the beta function, rather the coupling
itself, that can be explicitly determined. The arising of the scale parameter
 makes the interaction coupling 
change with the kinetic energy. 
Thus in different kinetic energy regions, distinct physical 
phenomena can be present even though they are dominated by the same
theory. One typical example is four-dimensional Quantum Chromodynamics (QCD), 
where at the high energy relative to the QCD scale parameter,  the quarks 
behave almost as free particles.  This asymptotic freedom 
property of quarks leads to the deep inelastic scattering
cross sections exhibiting a scaling behaviour.  At the low-energy 
level the quarks are confined and chiral symmetry breaking occurs.


However, in some cases  
spontaneous scale symmetry breaking is also possible. 
In a quantum field theory with no scalar field such as QCD and spinor
electrodynamics etc, spontaneous scale symmetry breaking usually 
takes place in 
the strong coupling region near a non-trivial 
fixed point of the beta function, 
and occurs in conjunction with spontaneous chiral 
symmetry breaking \cite{ref2}. It is well known that 
the dynamical breaking of chiral symmetry is characterized by the
fermion condensation $\langle \bar{\psi}\psi \rangle$, and that its occurrence
is determined by the composite operator effective potential generated by
quantum corrections, a function of $\langle \bar{\psi}(x)\psi(y)\rangle $
\cite{ref3}. At the fixed point,
the beta function vanishes and the theory will become scale invariant if
there are no dimensional parameters present in 
the classical theory \cite{ref4}. 
In particular, the running of the couplings freezes at the fixed point
and the anomalous scale symmetry 
breaking ceases to be a dominant effect. According to the arguments 
given in Ref.\cite{ref2}, if the dynamical chiral symmetry breaking occurs
when the anomalous breaking of scale symmetry is not dominant, 
then chiral symmetry breaking may imply the spontaneous breaking
of scale symmetry. It should be emphasized that the dynamical breaking
of chiral symmetry does not inevitably result in the spontaneous breaking
of scale symmetry, since the instability of the composite operator
effective potential under chiral symmetry does not necessarily 
lead to vacuum degeneracy  with respect to scale symmetry. 
In this case, the dynamical breaking
of chiral symmetry only results in the anomalous breaking of scale
symmetry. A method to identify spontaneous 
scale symmetry breaking is to observe whether there exists a  hierarchy
between the scale parameter, which governs the running of the coupling
constant and characterizes the anomalous breaking of scale symmetry, 
and the dynamically generated fermion mass. If there exists such 
a hierarchy, chiral symmetry
breaking can induce spontaneous scale symmetry breaking. Otherwise,
the dynamically generated fermion mass only leads to anomalous
scale symmetry breaking. 
The intuitive reason for this is that the scale parameter
characterizes the dimensional transmutation and the consequent
anomalous scale symmetry breaking. Thus the magnitude of the 
dynamically generated fermion mass associated with
the anomalous scale breaking should be of the same order
as the scale parameter, while if the dynamically generated fermion mass
has a hierarchy with the scale parameter, it should not attach to the 
anomalous breaking of scale symmetry and must result from
the spontaneous breaking of scale symmetry. A more rigorous but
less practical way to judge the spontaneous breaking of scale symmetry
is to check the fermion scattering amplitude to see
whether there exists a massless scalar particle called a dilaton, 
since according to Goldstone's theorem, there must arise a massless 
Goldstone particle with the same properties as the generator of the scale 
transformation as a consequence of the spontaneous breaking of scale 
symmetry.

Two typical examples were considered in \cite{ref2} to illustrate
this idea. The first example is four dimensional QCD.  Lattice
simulation indicates that the scale of chiral symmetry breaking
for fermions in higher dimensional representations of the 
gauge group is much higher than the confinement scale \cite{ref5}. 
This fact was further confirmed by checking the effective potential 
for the composite operator $\langle\bar{\psi}\psi\rangle$ together 
with the solution of the renormalization group equation \cite{ref6}. 
Thus there is spontaneous breaking of scale symmetry associated 
with fermion condensation. 
The other example is four-dimensional spinor electrodynamics 
at strong coupling. 
The occurrence of spontaneous scale symmetry breaking is due to
the existence of a non-trivial UV fixed point $\alpha_c$. 
The dynamically generated fermion mass
is proportional to the momentum scale \cite{ref2}, 
\begin{eqnarray}
B(0)~{\sim}~\Lambda \exp\left[-\frac{\pi}{\sqrt{\alpha/\alpha_c-1}}\right],
~~~~\alpha=\frac{e^2}{4\pi}.
\label{eq24}
\end{eqnarray}
However, according to Miransky's observation \cite{ref7}, the running
of the coupling constant can be written as 
\begin{eqnarray}
\frac{\alpha}{\alpha_c}=1+\frac{\pi^2}{\ln^2(\Lambda/\kappa)}
\label{eq25}
\end{eqnarray}
with $\kappa$ being an infrared cut-off. Eq.(\ref{eq25}) shows that
near the fixed point, $\alpha{\longrightarrow}\alpha_c$, 
the scale parameter $\Lambda$ must tend to infinity.  At the same time, 
it can be seen from Eq.(\ref{eq24}) that in the limit 
$\Lambda \rightarrow \infty$ the dynamically generated fermionic 
mass remains finite. This splitting produces a hierarchy 
between the fermion mass and the scale parameter. Therefore, near the 
UV fixed point of four-dimensional QED, the spontaneous breaking of 
scale symmetry occurs. Furthermore, 
it was explicitly shown that the scalar dilation manifests itself 
as the pole of the fermion-antifermion scattering amplitude \cite{ref2}.
 
Some time ago it was found that three-dimensional massless 
quantum electrodynamics (QED) can exhibit dynamically induced 
spontaneous chiral symmetry breaking \cite{ref8,ref9}. Compared 
with four dimensional quantum field 
theories, three dimensional QED has several special features. First,
it has an intrinsic dimensional parameter, the gauge coupling, that
plays the role of the scale parameter in four-dimensional QCD \cite{ref8}; 
Second, the beta function of three-dimensional quantum electrodynamics 
vanishes and hence the coupling constant stays fixed along
the whole trajectory of renormalization group flow. Thus 
there will arise no dimensional transmutation and consequently  
anomalous scale symmetry breaking does not happen perturbatively.
These facts seems to suggest that any spontaneous chiral 
symmetry breaking would induce spontaneous scale symmetry breaking.
However, it was argued implicitly in \cite{ref8} that 
there is no spontaneous breaking of 
scale symmetry at all in three-dimensional QED, 
despite the occurrence of dynamical chiral symmetry breaking. 
In this paper we intend to give a quantitative clarification of this point.

\section{ Scale Symmetry of 2+1-dimensional QED}

 As the first step, it is necessary to clarify the definition
of classical scale symmetry. The classical action of massless 
$2+1$-dimensional QED with $N_f$ flavours in the covariant Lorentz gauge 
is \cite{ref8}
\begin{eqnarray}
S&=&\int d^3 x {\cal L}=\int d^3x \left[\sum_{i=1}^{N_f}\bar{\psi}_i
\left(i\partial\hspace{-2.2mm}/ -e
A\hspace{-2.2mm}/\right){\psi}_i-\frac{1}{4}F_{\mu\nu}F^{\mu\nu}
-\frac{1}{2\xi}(\partial_\mu A^\mu)^2\right]. 
\end{eqnarray}
The coupling constant has the dimension 
$(\mbox{mass})^{1/2}$ 
and thus the theory is actually superrenormalizable. 
Further,  due to the vanishing beta function, the coupling constant
will remain frozen at quantum level.
The existence of such a coupling implies that the theory has 
an explicit scale symmetry breaking. This can be shown as
following. Under scale transformation, $x'=e^{-\epsilon}x$,  
the field transforms according to \cite{ref1}
\begin{eqnarray}
\phi'(x) = T(\epsilon)\phi(e^\epsilon x)\,;~~~~
T(\epsilon)=e^{\epsilon d_\phi}\nonumber
\end{eqnarray}
where  $\epsilon$ is the scale transformation parameter
and $d_\phi$ is the scale dimension of the field $\phi$. 
Thus the scale transformations for every fields are 
\begin{eqnarray}
\psi^{\prime}(x)=e^{\epsilon}\psi( e^{\epsilon} x), 
~~~A_{\mu}^{\prime}(x)=e^{1/2\epsilon}A_{\mu}( e^{\epsilon} x)
\end{eqnarray}
For an infinitesimal transformation we have
\begin{eqnarray}
&&\delta \psi_i=\epsilon (1+x^\mu\partial_\mu)\psi_i,~~~
\delta \left(\partial_{\mu}\psi_i\right)=\epsilon (2+x^\nu\partial_\nu)
\partial_\mu\psi_i,\nonumber\\
&&\delta A_\mu=\epsilon \left(\frac{1}{2}
+x^{\alpha}\partial_\alpha\right)A_{\mu}, ~~~
\delta \left(\partial_\nu A_{\mu}\right)=\epsilon 
\left(\frac{3}{2}+x^{\alpha}\partial_\alpha\right)\partial_{\nu}A_{\mu}, 
\label{eq5}
\end{eqnarray}
and consequently
\begin{eqnarray}
\delta S=\epsilon \int d^3x\left[\partial^{\mu}(x_\mu {\cal L})+
\frac{1}{2}e \sum_{i=1}^{N_f} \bar{\psi}_i A\hspace{-2.2mm}/ {\psi}_i\right].
\label{eq6}
\end{eqnarray}
The second term on the right hand side of Eq.(\ref{eq6}) is an explicit
violation of scale symmetry.
It can be shown, however, that the theory has an approximate 
scale invariance at scales for which 
the intrinsic energy scale $e^2 N_f$ can be ignored \cite{ref3}.
The breaking of scale
symmetry in Eq.(\ref{eq6}) is thus classical, and in the following 
we will examine whether or not quantum
corrections lead to a {\it dynamical} violation of the scale
symmetry.
 
The conserved quantity corresponding to above special scale
symmetry can be defined in the standard way.
The general variation of the classical action is 
\begin{eqnarray}
\delta S[\phi]
&=& \int d^3x \left[ \left(\frac{\partial {\cal L}}{\partial \phi}
-\partial_\mu \frac{\partial {\cal L}}{\partial (\partial_{\mu}\phi)}\right)
\delta \phi+\partial_{\mu}\left(\frac{\partial {\cal L}}
{\partial (\partial_{\mu}\phi)}\delta \phi \right) \right] .
\label{eq7}
\end{eqnarray}
For the scale transformation listed in Eq.(\ref{eq5}),
with the fields $\phi=(\bar{\psi}, \psi, A)$ satisfying 
the classical equations of
motion, Eqs.(\ref{eq5}) and (\ref{eq7}) yield
\begin{eqnarray}
\int d^3x\, \partial^{\mu}\left[d_{\phi}
\frac{\partial {\cal L}}{\partial (\partial^\mu\phi)}\phi+x_{\nu}
\left(\frac{\partial {\cal L}}{\partial (\partial^\mu\phi)}\partial_{\nu}\phi
-g_{\mu\nu}{\cal L}\right)\right]-
\frac{1}{2}e \sum_{i=1}^{N_f}\bar{\psi}_i A
\hspace{-2.2mm}/ {\psi}_i=0.
\label{eq8}
\end{eqnarray}
Defining the canonical energy-momentum tensor
 and the dilatation current in the same way 
as for the four dimensional scale invariant theory,
\begin{eqnarray}
\theta_{\mu\nu}^{(\rm can)}&=&\frac{\partial {\cal L}}
{\partial (\partial^\mu\phi)}\partial_{\nu}\phi-g_{\mu\nu}{\cal L}, \nonumber\\
d_{\mu}&=&d_{\phi}\frac{\partial {\cal L}}{\partial (\partial^\mu\phi)}\phi
+x^{\nu}\theta_{\mu\nu}^{(\rm can)},
 \end{eqnarray} 
we can write Eq.(\ref{eq8}) in the following form,
\begin{eqnarray}
\int d^3x \,\partial_{\mu}d^{\mu}=\int d^3x
\frac{1}{2}e \sum_{i=1}^{N_f}\bar{\psi}_i A
\hspace{-2.2mm}/{\psi}_i.
\end{eqnarray} 
An explicit calculation gives
\begin{eqnarray}
d_{\mu}=-\frac{1}{2}F_{\mu\nu}A^\nu-\frac{1}{2\xi}A_{\mu}
\partial_\nu A^\nu +x^\nu \theta_{\mu\nu}^{(\rm can)},
\end{eqnarray}
where
\begin{eqnarray}
\theta_{\mu\nu}^{(\rm can)}&=&\frac{i}{2}\sum_{i=1}^{N_f}\left(
\bar{\psi}_i\gamma_\mu \partial_\nu\psi_i-(\partial_\nu\bar{\psi}_i)
\gamma_\mu \psi_i\right)-F_{\mu\rho}\partial_\nu A^\rho-
\frac{1}{\xi}\partial_\nu A_\mu \partial_\rho A^\rho\nonumber\\
&&-\delta_{\mu\nu} \left[\frac{i}{2}\sum_{i=1}^{N_f}\left(
\bar{\psi}_i\gamma_\rho \partial^\rho\psi_i-(\partial^\rho\bar{\psi}_i)
\gamma_\rho \psi_i\right)-e\sum_{i=1}^{N_f}\bar{\psi}_iA\hspace{-2.2mm}/
{\psi}_i\right.\nonumber\\
&&\left.-\frac{1}{4}F_{\lambda\rho}F^{\lambda\rho}
-\frac{1}{2\xi}(\partial_\rho A^\rho)^2\right].
\end{eqnarray}
Using the classical equations of motion, 
\begin{eqnarray}
&& (i\partial\hspace{-2.2mm}/-eA\hspace{-2.2mm}/)\psi_i=0, ~~~~
i (\partial^\mu \bar{\psi}_i) \gamma_\mu+e \bar{\psi}_i A\hspace{-2.2mm}/=0,
\nonumber\\
&&\partial^{\nu}F_{\nu\mu}+\frac{1}{\xi}\partial_{\mu}
(\partial_\alpha A^\alpha)-\sum_{i=1}^{N_f} e \bar{\psi}_i \gamma_\mu
\psi_i=0,
\end{eqnarray}
we can easily verify
\begin{eqnarray}
&&\partial^\mu \theta_{\mu\nu}^{(\rm can)}=0, ~~~\theta^{({\rm can})\mu}_\mu
=-{\cal L}+\sum_{i=1}^{N_f}e \bar{\psi}_i A\hspace{-2.2mm}/\psi_i,\\
&&\partial_{\mu}d^{\mu}=\theta^{({\rm can})\mu}_\mu
+{\cal L}-\frac{1}{2}\sum_{i=1}^{N_f}e \bar{\psi}_i A\hspace{-2.2mm}/\psi_i
=\frac{1}{2}\sum_{i=1}^{N_f}e \bar{\psi}_i A\hspace{-2.2mm}/\psi_i.
\end{eqnarray}
When discussing anomalous scale symmetry 
breaking it is convenient to define an ``improved'' energy-momentum 
tensor $\theta_{\mu\nu}$ \cite{ref10} so that 
\begin{eqnarray}
d_\mu=x^{\nu}\theta_{\mu\nu}-\partial^\nu K_{\mu\nu}
\end{eqnarray}
with $\theta_{\mu\nu}=\theta_{\nu\mu}$ and $K_{\mu\nu}=-K_{\nu\mu}$.  
It is easy to show that 
\begin{eqnarray}
\partial^\mu \theta_{\mu\nu} =\partial^\mu \theta_{\mu\nu}^{(\rm can)}
=0, ~~~\theta^{\mu}_{~\mu}=\partial_{\mu}d^{\mu}
=\frac{1}{2}\sum_{i=1}^{N_f}e \bar{\psi}_i A\hspace{-2.2mm}/\psi_i.
\end{eqnarray}
The trace of the energy-momentum tensor will stay the same
at the quantum level since the beta function vanishes
and no trace anomaly or anomalous scale symmetry breaking arises \cite{ref4}. 
The explicit relation between the canonical and the improved 
energy momentum tensors is not so straightforward 
as in the case of four-dimensional 
scalar field theory \cite{ref10}. It will probably involve a nonlocal form 
of the fields \cite{ref11}.

The conserved charge corresponding to the scale symmetry is 
$D=\int d^3x d_0$
and the standard definition of spontaneous scale symmetry breaking is
given by $\widehat{D}|0\rangle =0$, where $\widehat D$ 
 denotes the corresponding
quantum operator for the dilatation generator. For the dynamical
spontaneous breaking of scale symmetry, we should
calculate the quantum effective potential composed of the expectation 
value of the composite operator $\bar{\psi}(x)\psi(y)$ \cite{ref3} 
and observe whether the fermionic mass coming from the instability of 
this effective potential under chiral symmetry has a hierarchy with the 
scale parameter \cite{ref2}. 

The above discussion has clearly shown the difference between scale 
symmetry and chiral symmetry. Scale transformation invariance is a space-time
symmetry, and is much more stringent than chiral symmetry 
in protecting the theory from receiving quantum corrections: 
quantum correction can explicitly break scale symmetry much more easily
than chiral symmetry. This is the 
reason that the spontaneous breaking of chiral symmetry does not
necessarily lead to the spontaneous breaking of scale symmetry,
despite the fact that their common feature is the dynamical generation
of a fermion mass term. Only when a further dynamical condition is 
satisfied (that 
the dynamical fermionic mass has a hierarchy with 
the scale parameter) does 
the dynamical breaking of chiral symmetry imply the spontaneous
breaking of scale symmetry. 

\section{Scale Symmetry Breaking at Zero Temperature}


To explore the dynamical breaking of chiral symmetry, we need to solve
the Schwinger-Dyson equation (SDE)
for the fermion self-energy
\begin{eqnarray}
 \Sigma (p)=-p\hspace{-2mm}/A(p)+B(p), ~~~~~p{\equiv}|p|.
\label{sde}\end{eqnarray}
The SDE for the fermion self-energy can be
obtained from extremizing the Cornwall-Jackiw-Tomboulis potential
with respect to $B(p)$. Thus, any nontrivial solution to the SDE
indicates the spontaneous breaking of chiral symmetry.  
However, it is notorious that the Schwinger-Dyson equations are 
a set of closed instantaneous integral equations which are impossible 
to solve completely. 
Some appropriate approximation must be utilized. The simplest choice
is the rainbow (ladder) approximation, which is consistent 
with the leading order of the large flavour number $N_f$ expansion 
\cite{ref8}. The Ward identity between the fermion self-energy
and the fermion-photon vertex requires that $A(p)=0$ under this approximation. 
In the rainbow approximation the gap equation 
in Landau gauge ($\xi=0$) reduces to \cite{ref8,ref9},
 \begin{eqnarray}
B(p)=\frac{e^2}{2\pi^2 p}\int_0^{\infty}dq\frac{q B(q)}{q^2+B^2(q)}
\ln\frac{p+q+N_fe^2/8}{|p-q|+N_fe^2/8}.
\label{eq19}
\end{eqnarray}
For $p{\ll}N_fe^2$ or  $p{\gg}N_fe^2$ the above integral 
equation can be converted into a second order nonlinear differential 
equation\cite{ref8}
 \begin{eqnarray}
\frac{d}{dp}\left[\frac{d B(p)}{dp}\frac{p^2(p+N_fe^2/8)^2}{2p+N_fe^2/8}
\right]=-\frac{N_fe^2}{\pi^2N_f}\frac{p^2B(p)}{p^2+B^2(p)}.
\label{eq191}
\end{eqnarray}
The fermion condensate  is given by
\begin{eqnarray}
 \langle\bar{\psi}(p)\psi(-p)\rangle {\sim} 
\mbox{Tr}[-p\hspace{-2mm}/+B(p)]^{-1}
=\frac{4B(p)}{p^2+B^2(p)}.
\end{eqnarray}
In the region $B(p){\ll}p{\ll}N_fe^2$ 
Eq.(\ref{eq191}) can be linearized and an analytical solution can be 
found. When $N_f<32/\pi^2$, the solution
has the form
\begin{eqnarray}
B(p)=\frac{(N_fe^2)^{3/2}}{\sqrt{p}}\sin\left[\frac{1}{2}\sqrt{
\frac{32}{N_f\pi^2}-1}\ln\left(\frac{p}{B(0)}\right)+\delta\right],
\label{eq22} 
\end{eqnarray}
where $\delta$ is a phase factor which may depend on the 
parameter $N_f$. The boundary conditions
\begin{eqnarray}
0{\leq} B(0)< \infty\,; ~~~\lim_{p{\rightarrow}N_fe^2}
\left[p\frac{dB(p)}{dp}+B(p)\right]=0
\end{eqnarray} 
together with Eq.(\ref{eq22}) determine that the dynamically generated 
mass defined at $p=0$ is \cite{ref8} 
\begin{eqnarray}
B(0){\sim}N_fe^2\exp\left[-\frac{2(n\pi-\delta)}{\sqrt{32/(\pi^2N_f)-1}}
\right].
\label{eq23}
\end{eqnarray}
This solution demonstrates the existence of a critical
flavour number $N_f^c=32/\pi^2$ for the restoration of chiral symmetry.

Despite some dispute about the existence of $N_f^c$ 
\cite{ref12},
a striking feature of (\ref{eq23}) is that the dynamically generated
mass is of the order of $N_fe^2$. In fact, this result is not
unexpected since no dimensional
transmutation occurs in $QED_3$ and thus there  is no 
dynamically generated momentum scale. The dynamical mass can only be  
associated  with the coupling constant, since the 
coupling constant is the only scale parameter of the theory.

In the region $p{\gg}N_fe^2$ Eq.(\ref{eq191}) cannot be linearized, but 
the asymptotic form of the solution has been obtained \cite{ref8},
\begin{eqnarray}
B(p){\sim}\frac{(N_fe^2)^3}{p^2}
\left[1-\left(\frac{1}{8}+\frac{2}{3N_f\pi^2}\right)
\frac{N_fe^2}{p}+{\cdots}\right],
\label{eq27}
\end{eqnarray}  
and the solution for the truncated lower-integral equation of 
Eq.(\ref{eq19}) is \cite{ref8}
\begin{eqnarray}
B(p){\sim}\frac{(N_fe^2)^{3}}{p(p+N_fe^2/8)}\left[\frac{p}{p+N_fe^2/8}
\right]^{8/(\pi^2N_f)}.
\label{eq28}
\end{eqnarray} 
The solution Eq.(\ref{eq28}) is only valid for $p{\gg}N_fe^2$
and becomes unreliable when $p{\sim}N_fe^2$. Eqs.(\ref{eq27})
and (\ref{eq28}) imply that $B(p){\sim}(N_fe^2)^3/p^2$. 
This result represents a dynamically generated mass function and thus 
leads to chiral symmetry breaking. Note that the solution 
(\ref{eq28}) has nothing to do with the $1/N_f$ expansion since 
this expansion does not play any role in this kinetic energy region.


When $p{\sim}N_fe^2$ it is difficult to get an analytical information
about the solution to the SDE (\ref{eq19}) and the only way to proceed 
is to use a numerical simulation. The numerical solution given in 
\cite{ref8} shows that when $p$ goes past $N_fe^2$ moving in the 
direction of increasing $p$, $B(p)$ sharply falls to zero.
In combination with the behaviour of the solutions to the SDE in the cases
 $p{\ll}N_fe^2$ and $p{\gg}N_fe^2$, one concludes that in the region
$p{\sim}N_fe^2$, $B(p)$ changes smoothly from its slowly falling
form at $p{\ll}N_fe^2$ to a sharply falling form at $p{\gg}N_fe^2$
\cite{ref8}.

Based on these results, we can observe whether or 
not a dynamical violation of scale symmetry occurs,
keeping in mind the discussion after Eq.~(\ref{eq6}) concerning
the this symmetry at the classical level. 
In the momentum range $p\ll N_fe^2$ Eq.(\ref{eq23}) indicates 
that the dynamical fermion mass is proportional to $N_fe^2$ and thus that
there is no hierarchy between the energy scale $N_fe^2$ and the 
dynamical mass $B(0)$. Therefore, it is not possible for a
dynamical violation of scale symmetry to take place. Note that if we approach 
the critical flavour number, the dynamical mass approaches zero and chiral 
symmetry breaking is lost, which means that there is still no possibility 
of dynamically broken scale symmetry. 
In the momentum range $p\gg N_fe^2$ 
we again have spontaneous chiral symmetry breaking since the dynamical 
mass function $B(p)$ behaves like $1/p^2$. 
However there exists no hierarchy between the dynamical mass
and the scale $N_fe^2$ in all cases and hence no dynamical violation
of scale symmetry breaking occurs.  



\section{Scale Symmetry Breaking at Finite Temperature}

Despite the fact that there is no spontaneous breaking of scale 
symmetry at zero temperature, it is possible that it could take
a place at finite temperature. An intuitive 
reason for this is as follows: temperature is a parameter 
with mass dimension and thus the dynamically generated mass
may depend not only on the coupling constant, but also on this new 
dimensional parameter. This new dimensional parameter 
makes it possible that
there will be a hierarchy between the dynamical mass and the
momentum scale $N_fe^2$ provided by the coupling constant. 
Moreover, the coupling constant at the quantum level may be
temperature dependent, and hence contrary to the zero temperature case,
there will arise a non-vanishing beta function which depends on the
temperature. If this beta function has a non-trivial fixed point,
then a similar observation as in the case of four-dimensional QED at
zero temperature could reveal the occurrence of the spontaneous 
breaking of scale symmetry.

As in the zero-temperature case, we make use of the Schwinger-Dyson
equation to observe the dynamical mass function.
The Schwinger-Dyson equation for the fermion self-energy 
at finite temperature $\Sigma (p_0,|{\bf p}|)$,
in  Landau gauge reads \cite{ref12}
\begin{eqnarray}
\Sigma (p_0, |{\bf p}|)&=&-p_0\gamma_0 A(p_0, |{\bf p}|)-{\bf p}{\cdot}
{\bf \gamma}A(p_0, |{\bf p}|)+B(p_0, |{\bf p}|)\nonumber\\
&=&-\frac{e}{\beta}\sum_{n=-\infty}^{\infty}\int\frac{d^2{\bf q}}{(2\pi)^2}
 \gamma_{\mu}S(q_0, |{\bf q}|)\Gamma_\nu
D^{\mu\nu}(p_0-q_0, |{\bf p}-{\bf q}|,\beta),
\label{eq29}
\end{eqnarray}
where $\beta=1/(k_B T)$, $p_0=(2 m+1)\pi/\beta$ and $q_0=(2n+1)\pi/\beta$.
To leading order in the $1/N_f$ expansion, the ladder approximation 
is appropriate: $\Gamma_{\mu}$ can be replaced by the bare
vertex $e\gamma_\mu$, and the propagator $D_{\mu\nu}(p_0,{\bf p},\beta)$ 
is approximately evaluated in the chain approximation of fermionic loop
\cite{ref12}.
The Ward identity further requires that $A(p_0, |{\bf p}|)=0$. The
dynamical mass function is obtained by taking the trace of Eq.(\ref{eq29})
\begin{eqnarray}
B(p_0, |{\bf p}|,\beta)=\frac{e^2}{\beta}\sum_{n=-\infty}^{\infty}\int
\frac{d^2{\bf q}}{(2\pi)^2}D(p_0-q_0,|{\bf p}-{\bf q}|,\beta)
\frac{B(q_0, |{\bf q}|,\beta)}{q^2+B^2(q_0, |{\bf q}|,\beta)}
\label{eq30}
\end{eqnarray}
where $D(p_0,|{\bf p}|,\beta){\equiv}\mbox{Tr}[\gamma_\mu 
D_{\mu\nu}(p_0,|{\bf p}|,\beta)\gamma_\nu]/8$ \cite{ref13}. The closed 
integral equation (\ref{eq30}) for $B(p_0, |{\bf p}|,\beta)$ has been 
numerically solved in an instantaneous exchange approximation in which 
the $p_0$ dependence of the vacuum polarization tensor has been ignored. 
As a consequence, $B(p_0, |{\bf p}|,\beta)$ also becomes 
frequency independent. 
In the approximation of considering only the $\mu=\nu=0$ component,
the propagator takes the form \cite{ref12},
\begin{eqnarray}
D_{\mu\nu}({\bf p-q},\beta)=\frac{\delta_{\mu 0}\delta_{\nu 0}}
{|{\bf p-q}|^2+\Pi (|{\bf p-q}|,\beta)},
\label{eq30a}
\end{eqnarray}
and 
\begin{eqnarray}
\Pi(|{\bf q}|,\beta)=\frac{2N_fe^2}{\pi\beta}\int_0^1 dx
\ln\left\{2\cosh\left[\beta/2 |{\bf q}| \sqrt{x (1-x)}\right]\right\}.
\label{eq32}
\end{eqnarray} 
After the summation over $n$ is performed, Eq.(\ref{eq30}) becomes \cite{ref13}
\begin{eqnarray}
B(|{\bf p}|,\beta)=\frac{e^2}{8\pi^2}\int d^2{\bf q}
\frac{B(|{\bf q}|,\beta)}{({\bf p}-{\bf q})^2+\Pi (|{\bf p}-{\bf q}|,\beta)}
\frac{\tanh \left[\beta/2\sqrt{{\bf q}^2+B^2(|{\bf q}|,\beta)}\,\right]}
{\sqrt{{\bf q}^2+B^2(|{\bf q}|,\beta)}]}.
\label{eq31}
\end{eqnarray}

The numerical solutions of Eq.(\ref{eq31}) have been explicitly obtained 
in the kinetic energy regions $|{\bf p}|<k_BT<N_fe^2$ 
and $k_BT< |{\bf p}| <N_fe^2$ \cite{ref13}. The analysis given below
will show that these are the only two regions in which dynamical 
chiral symmetry breaking 
takes place, thus they are the only regions that need to be considered. 
The numerical solution shows that there exists both a critical 
temperature and a critical flavour number, above which the dynamical
mass vanishes and chiral symmetry breaking is restored. The existence
of a critical temperature implies that we do not expect chiral 
or scale symmetry breaking at high temperature, and thus that it 
is sufficient to consider the range $k_BT< k_BT_c <N_fe^2$.

In the kinetic energy region
$|{\bf p}|<k_BT<N_fe^2$, the dynamical mass defined at $|{\bf p}|=0$ 
is a function of $N_fe^2$ and $k_BT$. Due to some special features
of the numerical calculation, the numerical solution can 
only describe the dependence of the mass on one of the two 
independent parameters  ($T$ and $N_f$) at a time.  
When we hold the temperature $T$ fixed, 
the numerical solution implies that the dynamical mass 
takes the following form \cite{ref13}
\begin{eqnarray}
B(0,\beta,e^2 N_f)\,{\propto}\,N_fe^2 
\exp\left[-\frac{C(T)}{\sqrt{N_c(T)/N_f-1}}\right],
\label{eq33}
\end{eqnarray}
where $C(T)$ is a certain temperature dependent function and $N_c(T)$ is
the critical flavour number (since $B(0,\beta,e^2 N_f){\rightarrow}0$ 
as $N_f{\rightarrow}N_c$). Eq.(\ref{eq33}) seems to suggest that 
no spontaneous breaking of scale symmetry is induced at 
any temperature since no matter how the temperature varies,
the dynamical mass is always of order $N_fe^2$. 
However, when we fix $N_f$ and look at the numerical 
solution for the dynamical mass as a function of $T$ we have \cite{ref13},
\begin{eqnarray}
B(0,\beta,e^2 N_f)\,{\propto}\,(N_fe^2)^{1-x(N_f)}
\left[k_B\left(T_c-T\right)\right]^{x(N_f)}; 
~~~~~k_BT{\sim}10^{-3}N_fe^2.
\label{eq34}
\end{eqnarray}
It has been shown that when $1<N_f<2$, the exponent $x$ has the value
$0.4<x(N_f)<0.6$ \cite{ref13}. 
Eq.(\ref{eq34}) not only explicitly shows the existence
of the critical temperature $T_c$, but also reveals that when 
$T{\rightarrow}T_c$ the dynamical mass $B(0,\beta,N_fe^2)$ is very small 
in comparison with the intrinsic energy scale $N_fe^2$.
Thus a gap between the dynamical mass and the energy scale $N_fe^2$
appears. This observation implies 
that spontaneous scale symmetry breaking occurs
in the region $|{\bf p}|<k_BT<N_fe^2$. Note also that, as stated above, 
the solution Eq.(\ref{eq34}) indicates the existence of a critical 
temperature $T_c$ and thus that at the high temperature  $T>T_c$ 
there will be no spontaneous breaking of chiral symmetry 
or of scale symmetry.

When $k_BT <|{\bf p}|\,{\leq}\,N_fe^2$, the dynamical mass cannot be defined
at $|{\bf p}|=0$.  For certain fixed temperatures, the numerical solution
shows that the ratio between the dynamical mass function and $N_fe^2$ 
decreases to zero when $|{\bf p}|\sim N_fe^2$ \cite{ref13} which indicates 
that there is no chiral symmetry breaking. When $k_BT<|{\bf p}|{\ll}N_fe^2$, 
if the flavour number is big enough,
the numerical solution indicates that $B(|{\bf p}|,\beta) \sim N_fe^2$.  
There is again no 
dynamical violation of scale symmetry. However,
when the flavour number is small, the numerical solution shows
that there exists a big gap between the dynamical mass and the scale 
$N_fe^2$, which suggests a dynamical violation
of scale symmetry. 

There is no information about the dynamical mass
in the the kinetic energy region $|{\bf p}|>N_fe^2>k_BT$. However, if
the dynamical mass function is a continuous and monotonically decreasing
function of $|{\bf p}|$, then the numerical solution at 
$|{\bf p}|{\sim}N_fe^2$ implies that the dynamical mass should
approach zero in this kinetic energy region. Therefore  
in this region chiral symmetry should be restored, and there
should be no dynamical violation of scale symmetry.

The Schwinger-Dyson equation (\ref{eq30})
has been  solved numerically  beyond the instantaneous exchange  
approximation, but the main features of the solutions 
remain unchanged \cite{ref14}. 
Therefore, the above analysis on the spontaneous
breaking of scale symmetry is likely to be valid beyond the instantaneous 
exchange approximation.

\section{Summary and Discussion}

Keeping in mind the discussion after Eq.~(\ref{eq6}) concerning
the nature of scale symmetry at the classical level, we have found
that the dynamical breaking of scale symmetry
is a very delicate non-perturbative phenomenon. Its occurrence is
not easy to identify since in most situations anomalous 
scale symmetry breaking prevails. In this paper
we have given a detailed analysis of spontaneous
scale symmetry breaking in $2+1$-dimensional QED based on solutions 
to the Schwinger-Dyson equations for the fermion self-energy 
at both zero and finite temperature.
In the case of zero temperature we show 
explicitly that scale symmetry breaking cannot be 
dynamically induced despite the fact that chiral symmetry breaking
occurs. The main reasons for this are the super-renormalizability
of the theory and the perturbative ultraviolet 
finiteness of $2+1$ dimensional QED.
These two facts eliminate the possibility for dimensional 
transmutation to occur 
and thus the only available scale parameter is the coupling
constant. Consequently, the dynamical mass must be proportional to 
the square of the coupling constant. The explicit solutions of the
SDE show that spontaneous breaking of scale symmetry, in the sense
of Eq.(\ref{eq6}) and the ensuing discussion, does not occur.

In the finite temperature case, the dynamical 
spontaneous breaking of chiral symmetry only
occurs in the kinetic energy region $|{\bf p}|<N_fe^2$, and 
only in the case $k_B T < N_f e^2$ with $k_BT<k_BT_c$ and $N_fe^2<N_ce^2$ 
($N_c$ and $T_c$ being the critical flavour number and critical temperature
to restore chiral symmetry).
When $|{\bf p}|<k_BT<N_fe^2$, the numerical solution to the
Schwinger-Dyson equation reveals that scale symmetry 
breaking may be induced dynamically, since there arises 
a hierarchy between the
dynamical mass and the energy scale $N_fe^2$.
When $k_BT<|{\bf p}|<N_fe^2$, in the case of small flavour number,
there is also an indication that spontaneous scale symmetry breaking
may take place.

It should be
emphasized, however, that the above conclusions about the phase structures 
of scale symmetry are based
on numerical solutions which are obtained by making some 
specific choices for the values of parameters and techniques used to solve 
the Schwinger--Dyson equation (\ref{sde}). Other,
more elaborate ans\"atzes, such as ones involving more complicated
vertex functions and subsequently non--trivial fermionic wave--function
renormalizations, can also be used, and provide a valuable
check on the consistency and completeness of these solutions \cite{ref14}. 
As such, 
these results should only be viewed as qualitative. A rigorous 
method to judge the spontaneous breaking of scale symmetry is
to calculate the fermion scattering amplitude
and observe whether it possesses a pole indicating the existence of 
the dilaton.

\section*{Acknowledgments}

We thank G.~Kunstatter for valuable discussions. W.F.C. 
also thanks Professors M. Chaichian
 for useful discussions
and encouragement.
This work is supported by the Natural Sciences and Engineering 
Research Council of Canada.

\end{document}